\documentclass[12pt]{iopart}
\usepackage{iopams}
\usepackage{graphicx}
\usepackage{subfig}
\usepackage{comment}
\expandafter\let\csname equation*\endcsname\relax
\expandafter\let\csname endequation*\endcsname\relax
\usepackage{amsmath}
\usepackage{amssymb}
\usepackage{dcolumn}
\usepackage{bm}
\usepackage{comment}
\usepackage{tikz}
\usepackage{multirow}
\usepackage{tabularx}
\usepackage{booktabs}
\usepackage{diagbox}

\begin{document}

\title{Revisiting the detection rate for axion haloscopes}

\author{Dongok Kim$^{1,2}$\footnotemark, Junu Jeong$^{1,2}$\footnotemark[\value{footnote}], \footnotetext{These authors contributed equally to this work.}, SungWoo Youn$^1$\footnotemark\footnotetext{Corresponding author.}, Younggeun Kim$^{1,2}$, Yannis K. Semertzidis$^{1,2}$}
\address{$^1$Center for Axion and Precision Physics Research, Institute for Basic Science, Daejeon 34051 Republic of Korea}
\address{$^2$Department of Physics, Korea Advanced Institute of Science and Technology, Daejeon 34141 Republic of Korea}
\ead{swyoun@ibs.re.kr}
\vspace{10pt}
\begin{indented}
\item[]\today
\date{\today}
\end{indented}

\begin{abstract}
The cavity haloscope has been employed to detect microwave photons resonantly converted from invisible cosmic axions under a strong magnetic field.
In this scheme, the axion-photon conversion power has been formulated to be valid for certain conditions, either $Q_{\rm cavity} \ll Q_{\rm axion}$ or $Q_{\rm cavity} \gg Q_{\rm axion}$.
This remedy, however, fails when these two quantities are comparable to each other.
Furthermore, the noise power flow has been treated independently of the impedance mismatch of the system, which could give rise to misleading estimates of the experimental sensitivity.
We revisit the analytical approaches to derive a general description of the signal and noise power.
We also optimize the coupling strength of a receiver to yield the maximal sensitivity for axion search experiments.
\end{abstract}

\vspace{2pc}

\section{Introduction}
The axion is the consequent product of the spontaneous breaking of the Peccei-Quinn global symmetry, and has been proposed as a dynamic solution to the CP problem in the strong interactions of particle physics~\cite{paper:axion}.
With some constraints on mass, the hypothetical particle possesses a cosomological implication which can account for dark matter in our galactic halo~\cite{paper:CDM}.
A large fraction of the mass range is accessible using the cavity haloscope technique where the axion field is resonantly converted into a microwave field in the presence of an external magnetic field~\cite{paper:sikivie}.
This detection scheme provides one of the most sensitive approaches to the well established theoretical models~\cite{paper:KSVZ, paper:DFSZ}.

The conversion power formula, which is broadly used by cavity haloscope experiments, is given by~\cite{paper:detect_rate}
\begin{equation}
P_{a\rightarrow\gamma\gamma} = g_{a\gamma\gamma}^2 \frac{\rho_a} {m_a}B_0^2VC\,{\rm min}(Q_c,Q_a),
\label{eq:conv_power_org}
\end{equation}
where $g_{a\gamma\gamma}$ is the axion-to-photon coupling, $\rho_a$ is the local halo axion density, $m_a$ is the axion mass, $B_0$ is the external magnetic field, $V$ is the cavity volume, $C$ is the form factor of the resonant mode under consideration, and $Q_c$ and $Q_a$ are the cavity and axion quality factors, respectively.
As a measure of experimental sensitivity, the signal-to-noise ratio (SNR) is conventionally adopted:
\begin{equation}
{\rm SNR} \equiv \frac{P_{\rm signal}}{\delta P_{\rm noise}},
\label{eq:snr}
\end{equation}
with $\delta P_{\rm noise}$ denoting the noise of the system.
The system noise is described by the Dicke radiometer equation as fluctuations in the noise power of the system within an integration time $\Delta t$ over a frequency bandwidth $\Delta\nu$~\cite{paper:radiometer}:
\begin{equation}
\delta P_{\rm noise}(\equiv \delta P_{\rm sys})=\frac{P_{\rm sys}}{\sqrt{\Delta t \Delta\nu}}=k_BT_{\rm sys}\sqrt{\frac{\Delta\nu}{\Delta t}},
\label{eq:noise_power_org}
\end{equation}
where the Johnson-Nyquist formula, $P_{\rm sys}=k_BT_{\rm sys}\Delta\nu$, is used with the Boltzmann constant $k_B$ and the equivalent system noise temperature $T_{\rm sys}$~\cite{paper:johnson-nyquist}.
Since the axion mass is a priori unknown, all possible mass ranges need to be explored.
In this regard, the figure of merit to be considered in an experimental system lies in how fast one can scan a mass region for a given sensitivity.
Using Eqs.~\ref{eq:conv_power_org} and~\ref{eq:noise_power_org}, Eq.~\ref{eq:snr} leads to the scanning rate in the form of 
\begin{equation}
\frac{df}{dt} =  g_{a\gamma\gamma}^4 \frac{\rho_a^2} {m_a^2} \frac{1}{\rm SNR^2} \frac{B_0^4V^2C^2}{k_B^2T_{\rm sys}^2} \frac{\beta^2}{(1+\beta)^2} Q_a {\rm min}(Q_l,Q_a),
\label{eq:scan_rate_org}
\end{equation}
where $\beta$ is the coupling strength of a receiver and $Q_l \equiv Q_c/(1+\beta)$ is the loaded cavity quality factor.
This formula gives rise to the optimal coupling strength $\beta=2$, which is typically chosen by many experiments to maximize sensitivity.

Current experiments~\cite{paper:ADMX}\cite{paper:HAYSTAC}\cite{paper:ORGAN}\cite{paper:CULTASK} have presumed so far that the last parameter in Eqs.~\ref{eq:conv_power_org} and~\ref{eq:scan_rate_org} always equals to $Q_c$ and $Q_l$ since the typical cavity quality factors of $Q_c\approx10^4\sim10^5$ are much smaller than the axion quality factor of $Q_a\approx10^6$~\cite{paper:axion_Q}.
However, the recent developments in superconducting (SC) technology and its applications to radio-frequency science are likely to dramatically increase the quality factor of resonant cavities even under strong magnetic fields~\cite{paper:sc_cavity}.
Furthermore, the system noise power in Eq.~\ref{eq:noise_power_org} has been assumed to be transmitted intact to the receiver regardless of impedance mismatch.
It is of importance to address that the noise power transmission is also subject to the receiver coupling in a similar manner to the signal power.
Here, we revisit the axion electrodynamics to derive a generic expression of the signal power, and consider a circuit diagram of an axion cavity haloscope to reformulate the system noise.
We also obtain the general form of the scanning rate and examine its dependency on the coupling strength to maximize experimental sensitivity.

\section{Axion electrodynamics}
In the presence of a coherently oscillating axion field $a(t)$, the Maxwell's equations in natural units are modified to~\cite{paper:electrodynamics}
\begin{equation*}
\begin{split}
&\nabla \cdot \left({{\bm E}-g_{a\gamma \gamma} a{\bm B}}\right)=\rho_{e},\\ 
&\nabla \cdot {\bm B}=0,\\
&\nabla\times{\bm E}=- \frac{\partial{\bm B}}{\partial t},\\
&\nabla\times \left({{\bm B}+g_{a\gamma \gamma}a{\bm E}}\right)=\frac{\partial}{\partial t} \left({{\bm E}-g_{a\gamma \gamma} a{\bm B}}\right)+{\bm J}_{e}.
\end{split}
\end{equation*}
The axion field is represented in the frequency domain by the Fourier transform as
\begin{equation*}
a(t) = \sqrt{T} \int_{-\infty}^{+\infty} \frac{d\omega}{2\pi} \mathcal{A}(\omega,\omega_a)e^{-i\omega t},
\end{equation*}
where $T$ is the time period over which the average is taken and $\omega_a$ is the axion angular frequency.
In the haloscope detection scheme, the axion field interacts with a static magnetic field, inducing an oscillating electromagnetic field that excites a mode of a microwave resonant cavity.
The axion-induced electric field has a general solution in the form of
\begin{equation*}
\begin{split}
{\bm E}_a({\bm r},\omega) &= g_{a\gamma \gamma}\mathcal{A}(\omega,\omega_a){\bm B}_0\left[1+\mathcal{F}(\omega,\omega_c)\mathcal{T}({\bm r}\omega)\right] \\
&\approx g_{a\gamma \gamma}\mathcal{A}(\omega,\omega_a){\bm B}_0 \mathcal{F}(\omega,\omega_c)\mathcal{T}({\bm r}\omega),
\end{split}
\end{equation*}
where $\mathcal{F}$ is the enhancement factor, typically $\mathcal{F}\gg1$, defined by the cavity resonant frequency $\omega_c$ and the cavity quality factor $Q_c$ as
\begin{equation*}
\mathcal{F}(\omega,\omega_c) = \frac{1}{(\omega-\omega_c)+i\omega_c/2Q_c}.
\end{equation*}
The function $\mathcal{T}$, determined by the cavity geometry, satisfies $\nabla^2 \mathcal{T}({\bm r}\omega)=\omega^2 \mathcal{T}({\bm r}\omega)$ and is related to the form factor as 
\begin{equation*}
\int_V \mathcal{T}^2({\bm r}\omega) d^3r = CV/\omega.
\end{equation*}

\section{Axion signal power}
The velocity distribution of the virialized dark matter axions in the galactic halo is assumed to have the Maxwell-Boltzmann structure, and thus the energy distribution of axion-induced photons in the Galaxy rest frame is represented by the Gamma distribution with a shape parameter of 3/2 and a scale parameter associated with the velocity dispersion. 
This distribution is then modified (boosted) by the circular motion of the solar system with respect to the Galaxy center and the orbital motion of the Earth around the Sun to yield the signal distribution that can be observed in the laboratory frame on Earth~\cite{paper:axion_Q}.
To a good approximation, the mean-square energy around the peak is described by the Cauchy function as~\cite{paper:hong}
\begin{equation*}
\begin{split}
\left\langle a^2(t)\right\rangle &= \int_{-\infty}^{+\infty}\frac{d\omega}{2\pi}\left|\mathcal A(\omega,\omega_a)\right|^2\\
&= \int_{-\infty}^{+\infty}\frac{d\omega}{2\pi}\left(\frac{4\rho_a Q_a}{\omega_am_a^2}\frac1{1+\left(\frac{\omega-\omega_a}{\omega_a/2Q_a}\right)^2}\right).
\end{split}
\end{equation*}
The validity of this approximation holds within less than 10\% (see Appendix).
Then the axion conversion power $P$ is written as
\begin{equation*}
\begin{split}
P&=\frac{\omega_cU}{Q_c}=\frac{\omega_c}{Q_c}\left(\int d^3r\left\langle\frac{{\bm E}_a^2+{\bm B}_a^2}2\right\rangle\right)\\
&=\frac{\omega_c}{Q_c}g_{a\gamma\gamma}^2B_0^2Q_c^2CV\int_{-\infty}^{\infty}\frac{d\omega}{2\pi} |\mathcal F(\omega,\omega_c)|^2 |\mathcal{A}(\omega,\omega_a)|^2,
\end{split}
\end{equation*}
where $\left\langle\frac{{\bm E}_a^2+{\bm B}_a^2}2\right\rangle = \left\langle E_a^2\right\rangle = \left\langle B_a^2\right\rangle$ holds on resonance~\cite{paper:eff_approx}.
As a result, the power is proportional to the product of two Cauchy functions
\begin{equation}
P\propto \int_{-\infty}^\infty\frac{d\omega}{2\pi}\left[\frac{1}{1+\left(\frac{\omega-\omega_c}{\omega_c/2Q_c}\right)^2}\right]\left[\frac1{{1}+\left(\frac{\omega-\omega_a}{\omega_a/2Q_a}\right)^2}\right].
\label{eq:power_cauchy}
\end{equation}

\begin{figure}[h]
\centering
\includegraphics[width=0.375\linewidth]{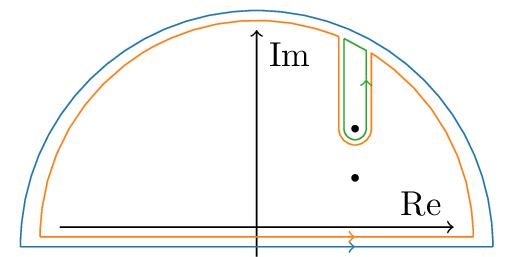}
\caption{Contour integrals in the complex plane. Two poles at resonance ($\omega_c=\omega_a$) are represented by the black dots.
The total integral (blue) is the sum of the sub-integrals represented by orange and green contours.}
\label{fig:cont_int}
\end{figure}

In the complex plane Eq.~\ref{eq:power_cauchy} has four poles at
\begin{equation*}
\omega = \omega_c \pm \frac{i}{2Q_c}\omega_c \,\, {\rm and} \,\, \omega = \omega_a \pm \frac{i}{2Q_a}\omega_a,
\end{equation*}
and the power is maximized when the peaks of the two functions coincide, i.e., when a resonance occurs at $\omega_c=\omega_a=\omega_0$.
This condition is represented by two poles in the upper half-plane, as shown in Fig.~\ref{fig:cont_int}, with the upper and lower poles corresponding to min($Q_c$,$Q_a$) and max($Q_c$,$Q_a$) respectively.

\begin{figure}[h]
\centering
\includegraphics[width=0.575\linewidth]{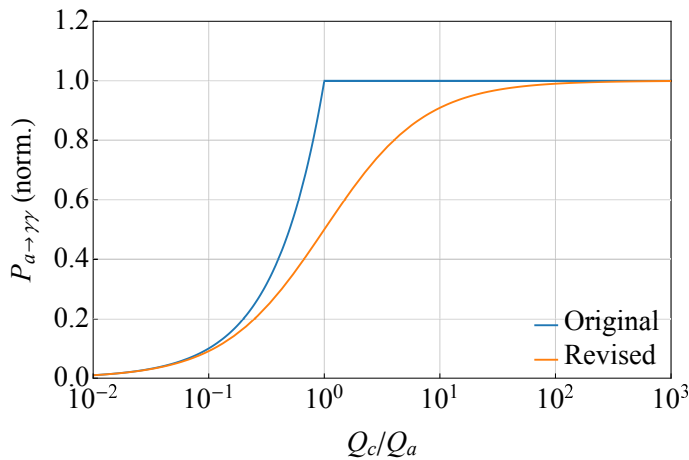}
\caption{Normalized conversion power as a function of $Q_c$ relative to $Q_a$. 
The original (Eq.~\ref{eq:conv_power_org}) and revised (Eq.~\ref{eq:conv_power_rev}) conversion powers are represented by blue and orange lines, respectively. 
They converge to each other at $Q_c/Q_a\rightarrow 0$ or $\infty$.}
\label{fig:conv_power}
\end{figure}

Then the integral can be decomposed into two sub integrals, as shown in Fig~\ref{fig:cont_int}, whose summation yields
\begin{equation*}
I=\frac{\omega_0/4}{\max(Q_c,Q_a)+\min(Q_c,Q_a)}=\frac{\omega_0/4}{Q_c+Q_a}.
\end{equation*}
From this, the general expression of the axion conversion power is obtained for arbitrary $Q_c$ as
\begin{equation}
P_{a\rightarrow\gamma\gamma}=g_{a\gamma\gamma}^2\frac{\rho_a}{m_a^2}B_0^2V\omega_0C\frac{Q_cQ_a}{Q_c+Q_a},
\label{eq:conv_power_rev}
\end{equation}
where we use $m_a=\omega_a$.
Eq.~\ref{eq:conv_power_rev} can be interpreted to mean that the conversion power is enhanced by the reduced quality factor of the system $Q_{\mu}$:
\begin{equation*}
\frac{1}{Q_{\mu}}=\frac{1}{Q_c}+\frac{1}{Q_a}.
\end{equation*}
For extreme cases, $Q_c \ll Q_a$ or $Q_c \gg Q_a$, Eq.~\ref{eq:conv_power_org} is recovered.
Figure~\ref{fig:conv_power} compares the conversion power estimated by the original (Eq.~\ref{eq:conv_power_org}) and revised (Eq.~\ref{eq:conv_power_rev}) equations as a function of the cavity quality factor.
It can be seen that the former contains a kink at $Q_c = Q_a$, while the latter exhibits smooth behavior.
It is also noticed that the original version overestimates the conversion power, particularly when the two quality factors are comparable.

Experimentally, the axion power generated in the cavity is extracted via an externally coupled receiver, e.g., a coaxial antenna.
Thus the detected signal power depends on the coupling strength of the receiver to the cavity and is given by
\begin{equation}
P_\textrm{signal}=\frac\beta{1+\beta}g_{a\gamma\gamma}^2\frac{\rho_a}{m_a}B_0^2VC\frac{Q_lQ_a}{Q_l+Q_a}.
\label{eq:sig_power_rev}
\end{equation}
The resonant enhancement factor, $Q_lQ_a/(Q_l+Q_a)$, corresponds to the total quality factor of the system including the receiver
\begin{equation*}
\frac{1}{Q_\mu}=\frac{1}{Q_c}+\frac{1}{Q_a}+\frac{1}{Q_r},
\end{equation*}
where $Q_r(=Q_c/\beta)$ is the effective quality factor of the receiver. 

\section{Noise power}
There are two major noise sources in cavity haloscope experiments: 1) Johnson-Nyquist thermal noise attributed to the blackbody radiation by the physical temperature of the cavity; and 2) noise added by the RF readout chain with its dominant contribution from the first stage amplifier.
Both noise components are dictated by the Dicke radiometer equation, in the same manner as Eq.~\ref{eq:noise_power_org}, as
\begin{equation}
\delta P_i = k_BT_i \sqrt{\frac{\Delta\nu_a}{\Delta t}},
\label{eq:noise_fluctuation}
\end{equation}
with $T_i$ denoting the equivalent temperature to either the thermal (physical) or added noise.
These two noise temperatures linearly contribute to the total system noise temperature, such that $T_{\rm sys} = T_{\rm phy} + T_{\rm add}$.

\begin{figure}[h]
\centering
\includegraphics[width=0.575\linewidth]{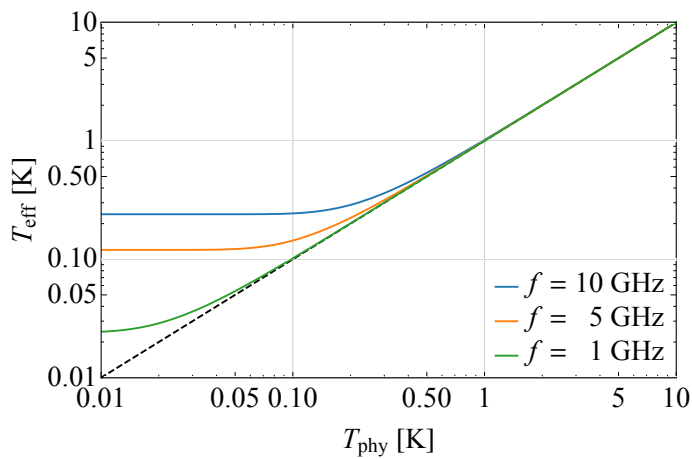}
\caption{Effective temperature, defined in Eq.~\ref{eq:eff_temp} vs. physical temperature for different frequencies: 1, 5 and 10\,GHz.
The dashed black line represents the classical approach.}
\label{fig:eff_temp}
\end{figure}

When estimating noise, it has been assumed so far that the thermal noise generated in the cavity is propagated to the receiver in its entirety without being subject to impedance mismatch between the two RF systems.
This may be because receiver chains are typically designed such that a circulator with one port terminated by a matched load at the end, which is equivalently an isolator, is placed between the cavity and the preamplifier.
The matched load appears to the amplifier regardless of the cavity-receiver coupling~\cite{thesis:rogers} and thus the loaded noise power on the amplifier is always the same as the generated one, $P_{\rm phy}=k_BT_{\rm phy}\Delta\nu$.
However, here we count the impedance mismatch in, to derive a general description of noise propagation.
In addition, with advanced cryogenic technologies, very low temperatures are achievable and thus quantum effects must be taken into account. 
According to Ref.~\cite{paper:johnson-nyquist}, the noise fluctuation at temperature $T_{\rm phy}$ for a given impedance $Z(\omega)$ of a system is 
\begin{equation*}
V_{\text{rms}}^{2} = 4 k_{B}T_{\rm phy}\eta(\omega) \text{Re}\left[ Z(\omega)\right],
\end{equation*}
where $V_{\rm rms}$ is the rms noise voltage generated in the cavity and the function $\eta(\omega)$ is in general given by
\begin{equation}
\eta(\omega) = \frac{\hbar\omega} {k_{B}T_{\rm phy}} \left( \frac{1}{e^{\hbar\omega/ k_{B}T_{\rm phy}} - 1}+\frac{1}{2} \right).
\label{eq:eta}
\end{equation}
The first term in the parenthesis in Eq.~\ref{eq:eta} is the average thermal photon number at frequency $\omega$ at temperature $T_{\rm phy}$, while the second term accounts for zero-point fluctuations, which becomes important in the quantum regime. 
We define the effective physical temperature 
\begin{equation}
T_{\rm eff} \equiv T_{\rm phy}\eta(\omega),
\label{eq:eff_temp}
\end{equation}
so that it reflects the quantum limit near absolute zero, as shown in Fig.~\ref{fig:eff_temp}. 

\begin{figure}[b]
\centering
\includegraphics[width=0.55\linewidth]{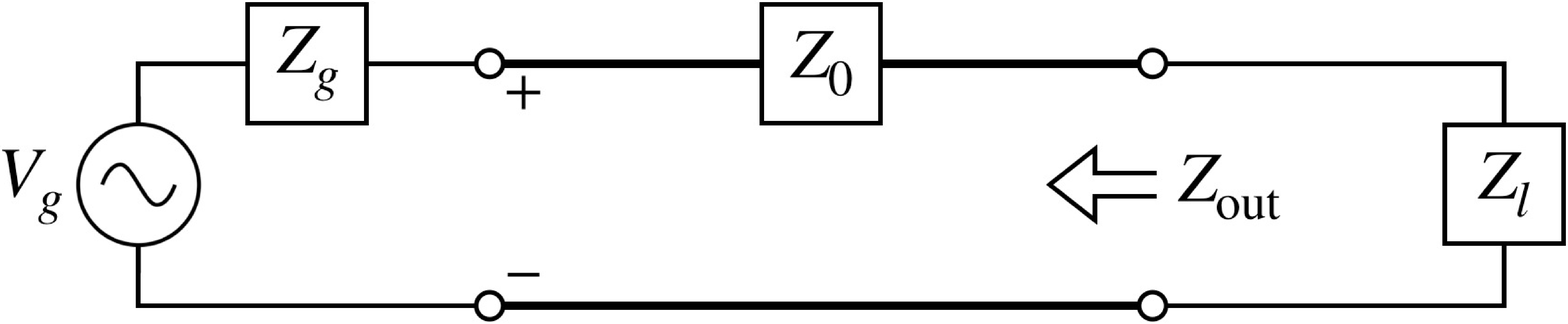}
\caption{Equivalent circuit diagram of a cavity haloscope experiment. $V_g$ is the voltage at the generator (cavity).
$Z_g$ and $Z_l$ are the impedances of the generator (loaded) and the load (detector).
The transmission line (receiver) has the characteristic impedance $Z_0$.
$Z_{\rm out}$ denotes the equivalent output impedance seen by the load.}
\label{fig:equiv_circuit}
\end{figure}

A typical axion haloscope consists of a resonant cavity connected to a detector chain via a receiver antenna.
This configuration can be treated as an equivalent RF circuit composed of a generator (cavity) and a load (detector) coupled by a transmission line (receiver) with respective impedances of $Z_g$, $Z_l$ and $Z_0$, as shown in Fig.~\ref{fig:equiv_circuit}.
The  average output power flow from the cavity to the detector is given by~\cite{book:pozar}
\begin{equation}
P_{\rm out}(\omega) = k_{B}T_{\rm eff} \Delta\nu_d \frac{4 \text{Re}\left[Z_l(\omega)\right] \text{Re}\left[ Z_{\rm out}(\omega)\right]}{\left|Z_l(\omega) + Z_{\rm out}(\omega) \right|^{2}},
\label{eq:P_out}
\end{equation}
where $\Delta \nu_d$ is the detector bandwidth and $Z_{\rm out}$ is the output impedance seen by the detector. 
For practical purposes, the transmission line is assumed to be lossless, so its characteristic impedance $Z_0$ is generally considered to be real.
And also the detector chain is designed to match the transmission line impedance such that $Z_l(\omega) \approx Z_0$, by which Eq.~\ref{eq:P_out} becomes
\begin{equation}
P_{\rm out}(\omega) \approx k_{B}T_{\rm eff} \Delta\nu_d \frac{4 Z_0 \text{Re}\left[ Z_{\rm out}(\omega)\right]}{\left|Z_0 + Z_{\rm out}(\omega) \right|^2}.
\label{eq:P_out_approx}
\end{equation}

For a transmission line with a length $l$ and propagation constant $\gamma$, the output impedance $Z_{\rm out}$ is given by the transmission line impedance equation as
\begin{equation}
Z_{\rm out}(\omega)=Z_0\frac{Z_g(\omega) + Z_0\tanh{\gamma l}}{Z_0 + Z_g(\omega)\tanh{\gamma l}}.
\label{eq:Z_out}
\end{equation}
Near the resonance region, $\omega \approx \omega_0$, $Z_g(\omega)$ is approximated to
\begin{equation}
\begin{split}
Z_{g}(\omega) &\approx Z_{g}(\omega_0) \left[1 + i\left(\frac{\omega-\omega_0}{\omega_0/2Q_l}\right) \right] \\
&= \frac{Z_{0}}{\beta} \left[1 + \frac{i}{1 + \beta} \left(\frac{\omega-\omega_0}{\omega_0/2Q_c}\right) \right].
\label{eq:Z_g}
\end{split}
\end{equation}
For a lossless transmission line, the propagation constant becomes purely imaginary, i.e., $\gamma=ik$, where $k$ is the wavenumber, and with Eq.~\ref{eq:Z_g}, Eq.~\ref{eq:Z_out} becomes
\begin{equation}
\begin{split}
Z_{\text{out}}(\omega)& = Z_{0}\frac{1 + i \beta \tan{k l}}{\beta + i \tan{k l}}, \\
& = Z_{0} \frac{2\beta +i(\beta^{2} - 1)\sin{2kl}}{(\beta^{2} + 1) + (\beta^{2} - 1)\cos{2kl}}.
\end{split}
\label{eq:Z_out_app}
\end{equation}

By plugging Eq.~\ref{eq:Z_out_app} into Eq.~\ref{eq:P_out_approx}, we compute the output power flow to the detector, i.e., the detected noise power, as
\begin{equation}
P_{\rm out}(\omega) = k_{B}T_{\rm eff} \Delta\nu_d \frac{4\beta}{(1 + \beta)^{2}},
\label{eq:power_out}
\end{equation}
which manifests its dependence on the coupling strength $\beta$.
Finally, the level of noise power fluctuations is determined by the radiometer equation to be
\begin{equation*}
\delta P_{\rm phy} = k_BT_{\rm eff} \frac{4\beta}{(1 + \beta)^2} \sqrt{\frac{\Delta\nu_a}{\Delta t}},
\end{equation*}
where we admire the fact that the SNR is maximal when the detector bandwidth equals the signal bandwidth~\cite{thesis:brubaker}.
It is noted that with $\beta=1$ and $\eta(\omega)=1$, Eq.~\ref{eq:noise_fluctuation} is recovered.

The added noise contribution from the receiver chain, on the other hand, is independent of $\beta$ and thus is simply given by
\begin{equation*}
\delta P_{\rm add} = k_BT_{\rm add} \sqrt{\frac{\Delta\nu_a}{\Delta t}},
\end{equation*}
with an equivalent noise temperature $T_{\rm add}$.
This noise is linearly added to the thermal noise to obtain the total system noise 
\begin{equation}
\begin{split}
\delta P_{\rm noise} &= \delta P_{\rm phy} + \delta P_{\rm add} \\
&= k_B \left(T_{\rm eff} \frac{4\beta}{(1 + \beta)^2} + T_{\rm add} \right) \sqrt{\frac{\Delta\nu_a}{\Delta t}},
\label{eq:noise_rev}
\end{split}
\end{equation}
where the sum of the two terms in the parenthesis is quoted as the system noise temperature.

An experimental demonstration was performed to verify the $\beta$ dependence of noise power transfer, as in Eq.~\ref{eq:power_out}, using a cylindrical copper cavity.
The cavity dimension provides a TM$_{010}$ resonant frequency of 5.9\,GHz and a cavity quality factor of 15,000 at room temperature.
The TM$_{010}$ mode is mostly chosen by experiments since it yields the maximum form factor for solenoids.
A monopole antenna is introduced on top of the cavity to pick up the power generated by the cavity thermal noise at room temperature.
The received noise power is amplified by a high-electron-mobility transistor before being delivered to a spectrum analyzer via a coaxial cable.
Based on the Y-factor method using a noise source with an excess-noise-ratio (ENR) of 15\,dB, the total gain and noise temperature of the detector chain were measured to be $36.7\pm0.3$\,dB and $75.2\pm3.1$\,K, respectively.
The power spectra from the cavity were acquired for various values of the coupling $\beta$, as measured by a network analyzer.
Different values of $\beta$ were obtained by varying the insertion length of the antenna into the cavity.
We also considered additional RF components, such as an isolator and a directional coupler, between the antenna and amplifier to examine their effects.
The resolution bandwidth of 10\,kHz was chosen because it was small enough, compared with the cavity width, to resolve signals, while also being large enough to achieve a reasonably fast acquisition time.

\begin{figure}[h]
\centering
\includegraphics[width=0.575\linewidth]{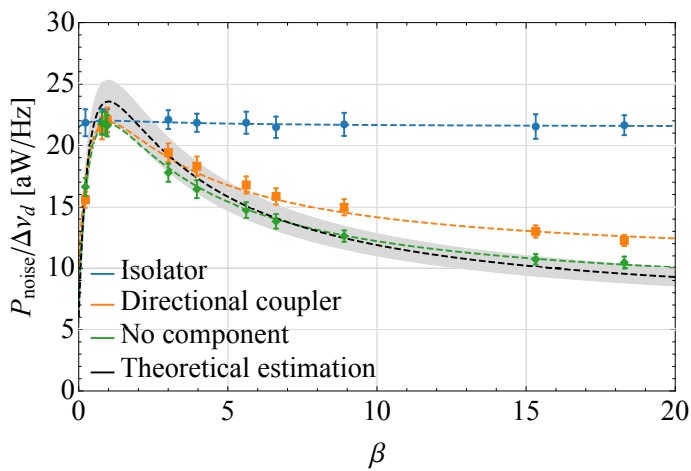}
\caption{Distributions of noise power at the resonant frequency of a cylindrical copper cavity as a function of coupling strength $\beta$.
A few RF components are considered between the cavity and the preamplifier - an isolator, directional coupler, and no component.
The data points, with statistical error bars, are fitted with a function $f(\beta)=A\cdot\beta/(1+\beta)^2+B$, which returns R-squared values of 0.025, 0.922, and 0.967, respectively.
The black dashed line is the expected power spectrum from the thermal noise at room temperature through a detector chain with a total gain of 36.7\,dB and noise temperature of 75.2\,K.
The gray band represents the uncertainty described in the text.}
\label{fig:noise_beta}
\end{figure}

Figure~\ref{fig:noise_beta} displays the noise power at the resonant frequency with various couplings for the different components.
The data points are fitted using a function of $\beta$ with the same form as in Eq.~\ref{eq:noise_rev}, i.e., $f(\beta) = A\cdot\beta/(1 + \beta)^2 + B$,
where $A$ and $B$ are the fitting parameters associated with $T_{\rm eff}$ and $T_{\rm add}$, respectively. 
R-squared was chosen as a statistical measure of the goodness of fit.
It is noticed that the fitting function describes the data reasonably well and small deviations from unity of the R-squared values imply the validity of the fit.
The estimated noise power distribution, assuming a total gain of 36.7\,dB and noise temperature of 75.2\,K at room temperature, is seen to be consistent with the measurement with no component.
The noise factor of the RF devices (here the amplifier) is known to have a dependence on the source impedance mismatch~\cite{proc:noise}, and its effect is also taken into consideration by comparing the noise power with the input port terminated with open, short, and matched loads.
The error band accounts for both the statistical uncertainties in the gain (0.3\,dB) and noise temperature (3.1\,K), and the systematic uncertainties from the inaccuracy of the ENR value (0.2\,dB) and the noise factor dependence on source impedance ($<4.6$\,K).

Insertion of the isolator makes the power transfer independent of the coupling strength, which confirms the aforementioned feature of isolators as matched loads.
The flat distribution is also statistically represented by a small R-squared value.
For the directional coupler, we find that the departure from the theoretical estimation particularly in the tail is attributed to the insertion loss of 0.2\,dB, which is consistent with the value obtained from the fit by taking a purely real propagation constant $\gamma$ into account in parameter $B$.
The maximum power appears when the impedance is matched ($\beta=1$) regardless of the type of added component and the corresponding values are comparable with each other.
Since an isolator brings higher noise to the detector at any coupling, it would be desirable to consider an alternative design for the detector chain in order to improve the SNR particularly for systems which require impedance mismatches.

\section{Experimental scanning rate}
The sensitivity of axion search experiments is determined by the SNR defined in Eq.~\ref{eq:snr}.
For convenience sake, we reformulate the signal power (Eq.~\ref{eq:sig_power_rev}) and system noise (Eq.~\ref{eq:noise_rev}) as
\begin{equation}
\begin{split}
P_\mathrm{signal} &= P_0\frac\beta{1+\beta}\frac{Q_lQ_a}{Q_l+Q_a} \,\,{\rm and}\\
\delta P_{\rm noise} &= k_BT_{\rm eff} \left( \frac{4\beta}{(1 + \beta)^2} + \lambda \right) \sqrt{\frac{\Delta\nu_a}{\Delta t}},
\label{eq:power_reform}
\end{split}
\end{equation}
with, respectively, $P_0\equiv g_{a\gamma\gamma}^2(\rho_a/m_a)B_0^2VC$ and $\lambda \equiv T_{\rm add}/T_{\rm eff}$, the relative contribution of the added noise with respect to the thermal noise of the system.
By putting Eq.~\ref{eq:power_reform} into Eq.~\ref{eq:snr}, the general form of the scanning rate is derived as 
\begin{equation}
\frac{df}{dt} =   \frac{1}{\rm SNR^2} \left(\frac{P_0}{k_BT_{\rm eff}}\right)^2 \left(\frac{\frac\beta{(1+\beta)}}{\frac{4\beta}{(1+\beta)^2}+\lambda}\right)^2 \frac{Q_lQ_a^2}{Q_l+Q_a}.
\label{eq:scan_rate_rev}
\end{equation}

For a fixed $Q_a$, the coupling $\beta$ can be optimized to maximize the detection rate for a given cavity quality factor $Q_c$ and noise contribution ratio $\lambda$.
The general solution for the optimal coupling, $\beta_{\rm opt}$, is obtained by requiring $\frac{d}{d\beta}\left(\frac{df}{dt}\right) =0$, which arrives at
\begin{equation}
\begin{split}
-\lambda\beta^4-(\lambda-4)\beta^3&+(8\tilde{Q}+2\lambda \tilde{Q}+\lambda-4)\beta^2\\
&+(4\lambda \tilde{Q}+\lambda)\beta+2\lambda \tilde{Q}=0,
\end{split}
\label{eq:beta_opt}
\end{equation}
where we use $\tilde{Q} \equiv Q_c/Q_a + 1$ for simplicity.
Four non-trivial solutions exist which satisfy Eq.~\ref{eq:beta_opt}, but in physically meaningful conditions of $\lambda\geq0$ and $\tilde{Q}\geq1$, there is only one real positive solution.
If the added noise is the dominant contribution to the total noise, i.e., $\lambda\gg1$, the optimal $\beta$ for a given $Q_c$ is obtained to be
\begin{equation*}
\label{eq:opt_beta}
\beta_{\rm opt}=\frac12\left(1+\sqrt{9+8Q_c/Q_a}\right).
\end{equation*}
It is noted that for $Q_c\ll Q_a$, the scanning rate and the optimal coupling become respectively
\begin{equation*}
\frac{df}{dt}\propto\frac{\beta^2}{(1+\beta)^2}Q_aQ_l\;\;{\rm and}\;\;\beta_{\rm opt}=2,
\end{equation*}
which restores the descriptions in Eq.~\ref{eq:scan_rate_org}.

In particular, we consider various scenarios of experimental design with practically interesting parameter values for $Q_c$ and $\lambda$: $Q_c/Q_a=10^{-2}-10^{2}$, reflecting the present situation and potential application of the SC cavity technology, and $\lambda=[10,\,1,\,0.1]$, representing transistor-based amplification, standard quantum-limited noise, and quantum noise in squeezed states, respectively.
A recently proposed experiment~\cite{paper:KLASH} could also belong to the third category ($\lambda=0.1$).
Tables~\ref{tab:beta_opt} and ~\ref{tab:dfdt_opt} list the optimal coupling strength and the corresponding scanning rate (normalized) for possible combinations of these scenarios.
\begin{table}[h]
\centering
\caption{Optimized coupling strength, $\beta_\mathrm{opt}$, obtained from Eq.~\ref{eq:beta_opt}, for various cavity quality factors and relative noise contributions.}
\label{tab:beta_opt}
\begin{indented}
\item[]
\begin{tabular}{c|rrr}
\br
\backslashbox{$Q_c/Q_a$}{$\lambda$}& ~~~~~~~~~~~10 &~~~~~~~~~~~~1 &~~~~~~~~~~ 0.1 \\
\mr
$10^{-2}$ & 2.2 & 4.7 & 40.1\\
$10^{-1}$ & 2.3 & 4.9 & 40.3\\
$10^0$ & 2.9 & 6.1 & 42.0\\
$10^1$ & 6.0 & 12.1 & 54.8\\
$10^2$ & 17.2 & 33.5 & 112.4\\
\br
\end{tabular}
\end{indented}
\end{table}

\begin{table}[h]
\caption{Maximum scanning rate computed by Eq.~\ref{eq:scan_rate_rev} using $\beta_{\rm opt}$ in Table~\ref{tab:beta_opt}.
The values are normalized to that for $Q_c/Q_a=10^{-2}$ and $\lambda=1$, which represent the currently achievable experimental parameters.}
\label{tab:dfdt_opt}
\begin{indented}
\item[]
\begin{tabular}{c|rrr}
\br
\backslashbox{$Q_c/Q_a$}{$\lambda$}& ~~~~~~~~~~~10 &~~~~~~~~~~~~1 &~~~~~~~~~~ 0.1 \\
\mr
$10^{-2}$ & $<0.1$ & 1 & 12\\
$10^{-1}$ & 0.3 & 10 & 127\\
$10^0$ & 2.0 & 87 & 1245\\
$10^1$ & 8.2 & 470 & 10565\\
$10^2$ & 15.2 & 1185 & 52898\\
\br
\end{tabular}
\end{indented}
\end{table}

\begin{figure}[b]
\centering
\includegraphics[width=0.575\linewidth]{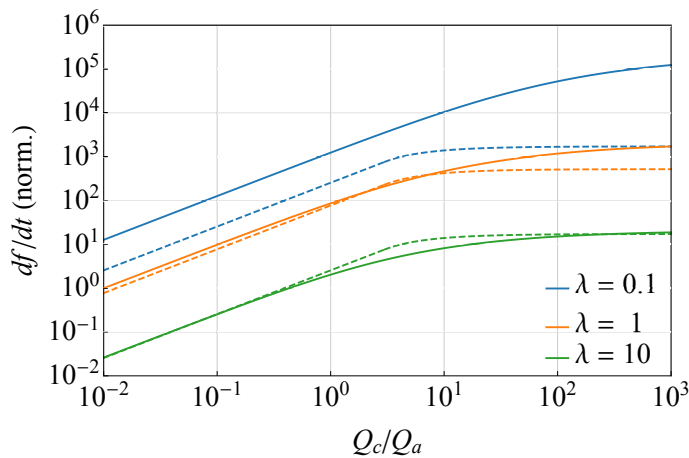}
\caption{Comparison of the scanning rate between the original (Eq.~\ref{eq:scan_rate_org}) and revised (Eq.~\ref{eq:scan_rate_rev}) calculations as a function of normalized cavity quality factor, $Q_c/Q_a$, for three different values of $\lambda$, the relative noise contribution. The former and the latter estimations are represented by dashed and solid lines, respectively.}
\label{fig:scan_rate_comp}
\end{figure}

The revised description on scanning rate (Eq.~\ref{eq:scan_rate_rev}) is compared with the original counterpart (Eq.~\ref{eq:scan_rate_org}) in Fig.~\ref{fig:scan_rate_comp} in terms of its dependency upon $Q_c$ and $\lambda$.
For the original version, we use $\eta(\omega)=1$ and $\beta_{\rm opt}=2$ to restore the classical approaches.
It it seen that 1) the expected detection rate gradually grows with increasing $Q_c$ in the high $Q_c$ region rather than being flattened out, and 2) there is a stronger dependence on $\lambda$ indicating substantial improvements are conceivable with less added noise contribution.
For current experiments operating with $Q_c/Q_a\sim10^{-2}$ and $\lambda \sim 1$, the sensitivity could have been slightly underestimated.
However, it is inferred that future experiments will take more advantage of further developments in superconducting RF cavities and quantum technology for higher sensitivities.

\section{Conclusion}
For cavity haloscope axion search experiments, sensitivity has been estimated based on limited assumptions, such as $Q_c\ll Q_a$ and noise power transfer free from impedance mismatch.
In this study, we extend those assumptions, by reflecting potential improvements in the cavity quality factor and taking into account the impedance mismatch in noise flow, to derive more comprehensive expressions for both signal and noise power.
We find the detection rate is sensitive to the receiver coupling strength, which was optimized for various scenarios of the experimental parameters.
A comparison with the original calculation indicates that further development of superconducting RF science and quantum technology would be more beneficial than we expected.
This revision would have non-trivial impacts on not only the sensitivity calculation of present experiments but also the conceptual design of future experiments.

\ack{This work was supported by IBS-R017-D1-2020-a00 / IBS-R017-Y1-2020-a00.}

\appendix
\section{Validity of the Cauchy approximation}
\label{sec:cauchy}
The energy spectrum of halo axions in the galactic rest frame is thermodynamically modeled by a Gamma distribution with a shape parameter of 3/2 and a scale parameter associated with the energy dispersion.
The spectrum, however, appears boosted by an observer on Earth owing to the Sun's circular motion as described in Ref.~\cite{paper:axion_Q} and the analytical form is rephrased as
\begin{equation}
\tilde{a}(u) = \sqrt{\frac{3}{2\pi}}\frac{1}{r} \exp{\left[-\frac{3}{2}\left(r^{2}+u\right) \right]}\sinh{\left[3r\sqrt{u}\right]},
\label{eq:exact}
\end{equation}
where $u$ is the axion kinetic energy in the unit of $m_{a}v_{\rm rms}^{2}/2$ with its mass $m_{a}$ and rms velocity $v_{\rm rms} \approx 270$\,km/s, and $r=v_{\odot}/v_{\rm rms} \approx 0.85$ is the boosting ratio with $v_{\odot}$ being the circular velocity of the Sun.
Ignoring the orbital and rotational motions of the Earth and using the average ($E_a=m_a\bar{v}_a^2/2$) and dispersion ($\Delta E_a=m_a\bar{v}_{\rm dis}^2/2$) of the kinetic energy, the boosted distribution function returns the average axion quality factor of $1.0\times10^6$. 

\begin{figure}[h]
\centering
\subfloat[]{\includegraphics[width=0.47\linewidth]{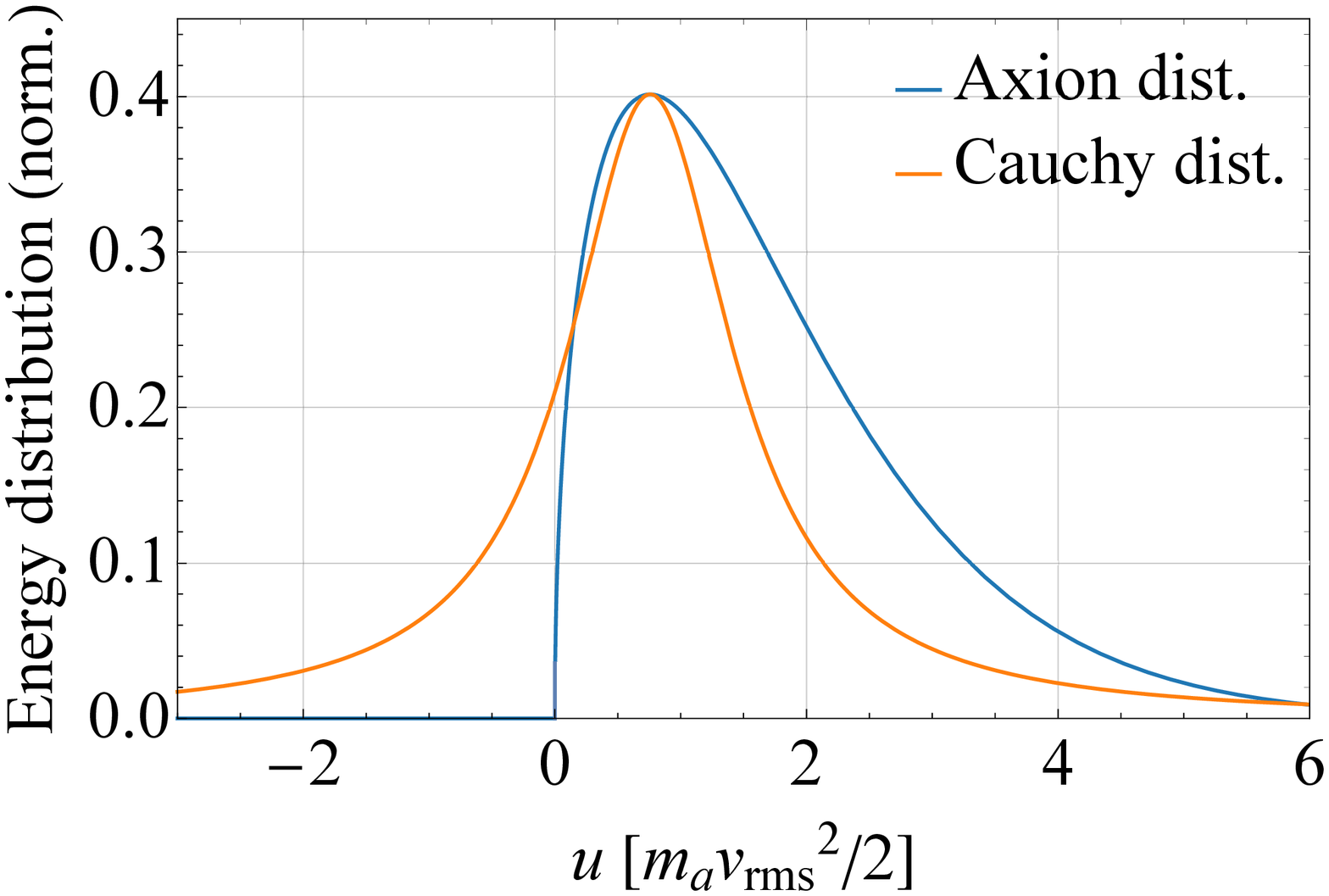}}
\hspace{0.02\linewidth}
\subfloat[]{\includegraphics[width=0.48\linewidth]{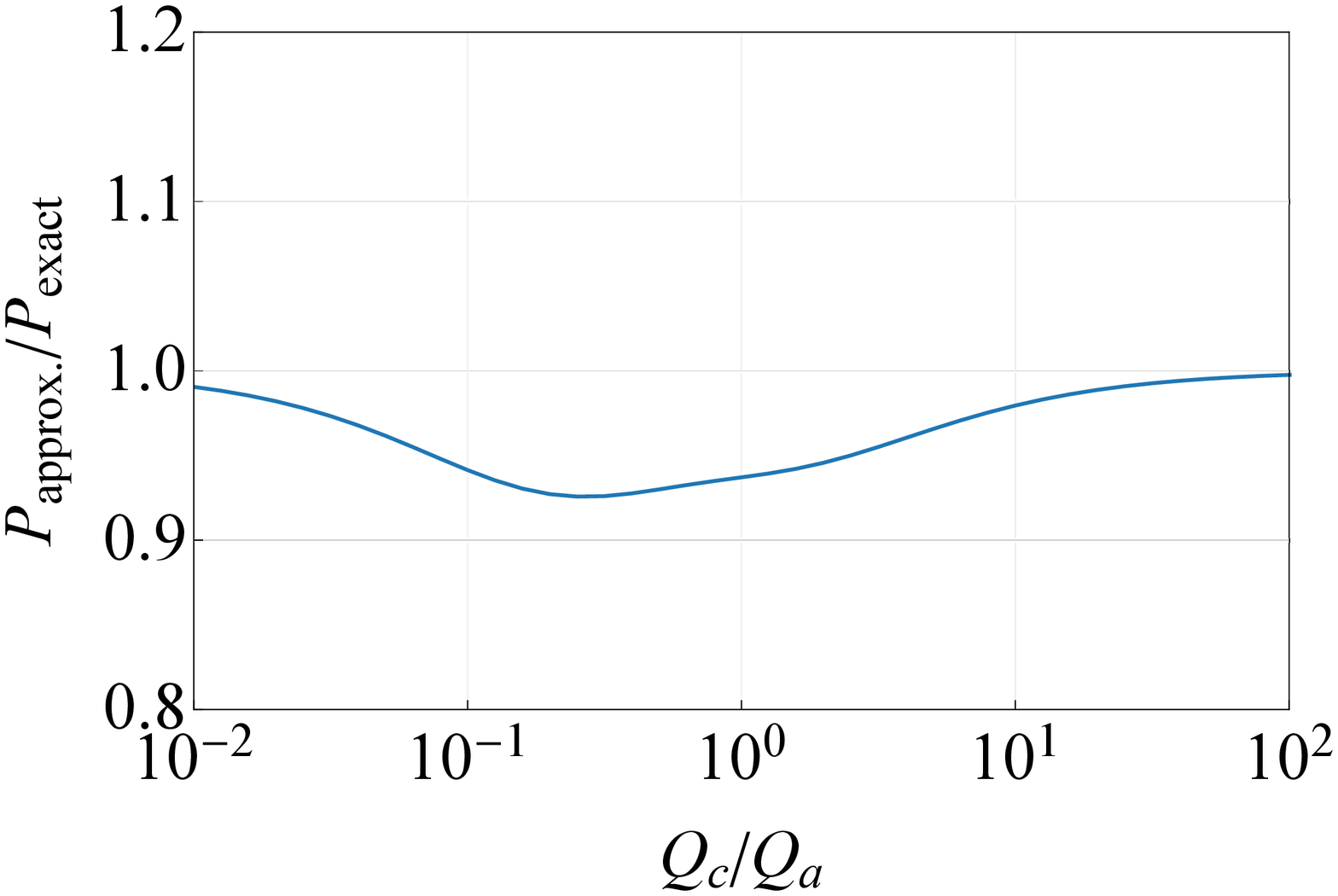}}
\caption{(a) Comparison of the axion energy distributions between the exact description from Eq.~\ref{eq:exact} and its Cauchy approximation in Eq.~\ref{eq:cauchy} in terms of the axion kinetic energy~$u$. 
The area below each distribution is normalized to unity and the peaks are matched.
(b) Ratio between the two integrals in Eq.~\ref{eq:integral} as a function of $Q_c/Q_a$.}
\label{fig:approx}
\end{figure}

For cavity haloscopes, the axion conversion power is proportional to the inner product of the axion distribution function and the cavity transfer function, which is represented by the Cauchy distribution.
An approximation of the axion energy spectrum to the Cauchy distribution,
\begin{equation}
\mathcal{C}_{\omega_a}^{Q_a}(\omega) = \frac{4Q_a}{\omega_a} \frac{1}{1+4Q_a^2(\omega/\omega_a-1)^2}
\label{eq:cauchy}
\end{equation}
with the axion frequency $\omega_a$ and quality factor $Q_a$, enables analytical calculations of the product not only near resonance but also over the entire spectrum.
Figure~\ref{fig:approx}(a) compares the exact axion distribution from Eq.~\ref{eq:exact} with the Cauchy approximation following Eq.~\ref{eq:cauchy}.
The area below each distribution is normalized to unity to have an equal amount of integrated power over the spectrum, and the peaks are aligned with the same height to have equal power on resonance.
These requirements determine the width of the Cauchy distribution, which defines a new quality factor that can be interpreted as the effective axion quality factor to be observed by a cavity haloscope.

The following integrals represent the conversion power in the exact and approximated approaches for the axion distribution: 
\begin{equation}
\int_{-\infty}^{\infty} \tilde{a}(\omega) C_{\omega_{a}}^{Q_{c}}(\omega) \frac{d\omega}{2\pi} \,\,{\rm vs.} \int_{-\infty}^{\infty} C_{\omega_{a}}^{Q_{a}}(\omega) C_{\omega_{a}}^{Q_{c}}(\omega) \frac{d\omega}{2\pi}.
\label{eq:integral}
\end{equation}
The numerical calculations were performed and their ratio as a function of $Q_c/Q_a$ is shown in Fig.~\ref{fig:approx}(b).
The Cauchy approximation underestimates the axion power by up to 8\% where $Q_c$ is comparable with $Q_a$, while it converges to the exact description at both extrema.
The effective axion frequency and quality factor observed by the haloscope are given, respectively, by
\begin{equation*}
\begin{split}
\omega_{a} &= m_{a}\left[1 + 0.77 \times \frac{1}{2}\left(\frac{v_{\rm rms}}{c}\right)^{2}\right] \approx m_a\,\,{\rm and} \\
\qquad Q_{a} &= \frac{\omega_{a}}{2/(\pi \tilde{a}(\omega_{a}))} \approx 1.6 \times 10^{6}.
\end{split}
\end{equation*}

Lastly, here we limit our work to the standard halo model where axion dark matter is isothermally distributed~\cite{paper:SHM}. 
In fact, more complicated velocity profiles are also possible for axions in different scenarios such as axion miniclusters~\cite{paper:miniclusters}, Bose stars~\cite{paper:bose_stars} and axion streams~\cite{paper:caustic_rings}, which could affect the signal observed by cavity haloscope experiments and thus require further studies. 



\section*{References}
\begin{thebibliography}{99}
\bibitem{paper:axion} R. D. Peccei and H. R. Quinn, {\it CP conservation in the presence of pseudoparticles}, Phys. Rev. Lett. {\bf 38}, 1440 (1977); 
S. Weinberg, {\it A new light boson?}, Phys. Rev. Lett. {\bf 40}, 223 (1978); 
F. Wilczek, {\it Problem of strong P and T invariance in the presence of instantons}, Phys. Rev. Lett. {\bf 40}, 279 (1978).
\bibitem{paper:CDM} J. Preskill, M. B. Wise and F. Wilczek, {\it Cosmology of the invisible axion}, Phys. Lett. B {\bf 120}, 127 (1983); 
L. F. Abbott and P. Sikivie, {\it A cosmological bound on the invisible axion}, Phys. Lett. B {\bf 120}, 133 (1983); 
M. Dine and W. Fischler, {\it The not-so-harmless axion}, Phys. Lett. B {\bf 120}, 137 (1983).
\bibitem{paper:sikivie}  P. Sikivie, {\it Experimental tests of the ``invisible" axion}, Phys. Rev. Lett. {\bf 51}, 1415 (1983).
\bibitem{paper:KSVZ} J. E. Kim, {\it Weak-interaction singlet and strong CP invariance}, Phys. Rev. Lett. {\bf 43}, 103 (1979); M. A. Shifman, A. I. Vainshtein, and V. I. Zakharov, {\it Can confinement ensure natural CP invariance of strong interactions?}, Nucl. Phys. B {\bf 166}, 4933 (1980).
\bibitem{paper:DFSZ} A. P. Zhitnitsky, {\it On the possible suppression of axion hadron interactions}, Yad. Fiz. {\bf 31}, 497 (1980) [Sov. J. Nucl. Phys. {\bf 31}, 260 (1980)]; M. Dine, W. Fischler and M. Srednicki, {\it A simple solution to the strong CP problem with a harmless axion}, Phys. Lett. B {\bf 104}, 199 (1981).
\bibitem{paper:detect_rate} P. Sikivie, {\it Detection rates for ``invisible"-axion searches}, Phys. Rev. D {\bf 32}, 2988 (1985).
\bibitem{paper:radiometer} R. H. Dicke, {\it The Measurement of Thermal Radiation at Microwave Frequencies}, Rev. Sci. Instr. {\bf 17}, 268 (1946).
\bibitem{paper:johnson-nyquist} J. B. Johnson, {\it Thermal agitation of electricity in conductors}, Phys. Rev. {\bf 32}, 97 (1928); H. Nyquist, {\it Thermal agitation of electric charge in conductors}, Phys. Rev. {\bf 32}, 110 (1928).
\bibitem{paper:ADMX} N. Du {\it et al.}, (ADMX Collaboration), {\it A Search for Invisible Axion Dark Matter with the Axion Dark Matter Experiment}, Phys. Rev. Lett. {\bf 120}, 151301 (2018).
\bibitem{paper:HAYSTAC} L. Zhong {\it et al.}, (HAYSTAC Collaboration), {\it Results from phase 1 of the HAYSTAC microwave cavity axion experiment}, Phys. Rev. D {\bf 97}, 092001 (2018).
\bibitem{paper:ORGAN} B. T. McAllister {\it et al.}, {\it The ORGAN Experiment: An axion haloscope above 15 GHz}, Phys. Dark Univ. {\bf 18}, 67 (2017).
\bibitem{paper:CULTASK} Y. K. Semertzidis {\it et al.}, {\it Axion Dark Matter Research with IBS/CAPP}, arXiv:1910.11591 (2019).
\bibitem{paper:axion_Q} M. S. Turner, {\it Periodic signatures for the detection of cosmic axions}, Phys. Rev. D {\bf 42}, 3572 (1990).
\bibitem{paper:sc_cavity} 
D. Alesini  {\it et al.}, {\it Galactic axions search with a superconducting resonant cavity}, Phys. Rev. D {\bf 99}, 101101 (2019);
D. Ahn {\it et al.}, {\it Maintaining high Q-factor of superconducting YBa$_2$Cu$_3$O$_{7-x}$ microwave cavity in a high magnetic field}, arXiv:1904.05111 (2019). 
\bibitem{paper:electrodynamics} F. Wilczek, {\it Two Application of Axion Electrodynamics}, Phys. Rev. D {\bf 58}, 1799 (1987).
\bibitem{paper:hong} J. Hong, J. E. Kim, S. Nam, and Y. K. Semertzidis, {\it Calculations of resonance enhancement factor in axion-search tube-experiments}, arXiv:1403.1576 (2014).
\bibitem{paper:eff_approx} Y. Kim {\it et al.}, {\it Effective approximation of electromagnetism for axion haloscope searches}, Phys. Dark Univ. {\bf 26}, 100362 (2019).
\bibitem{thesis:rogers} J. T. Rogers, {\it Limits on the electromagnetic coupling and density of galactic axions}, Ph.D. Thesis, University of Rochester, New York (1987).
\bibitem{book:pozar} D. M. Pozar, {\it Microwave Engineering \rm{(4th ed.)}}, John Wiley \& Sons, Inc. (2011).
\bibitem{thesis:brubaker} B. M. Brubaker, {\it First results from the HAYSTAC axion search}, Ph.D. Thesis, Yale University, New Haven (2018).
\bibitem{proc:noise} W. A. Haus {\it et al.}, {\it Representation of noise in linear twoports}, Proc. IEEE {\bf 48}, 69 (1960).
\bibitem{paper:KLASH} D. Alesini {\it et al.}, {\it KLASH Conceptual Design Report}, arXiv:1911.02427 (2019).
\bibitem{paper:SHM} A. Drukier, K. Freese, and D. Spergel, {\it Detecting cold dark matter candidates}, Phys. Rev. D {\bf 33}, 3495 (1986).
\bibitem{paper:miniclusters} C.J. Hogan and M.J. Rees, {\it Axion miniclusters}, Phys. Lett. B, {\bf 205}, 228 (1988).
\bibitem{paper:bose_stars} E.W. Kolb and I. I. Tkachev, {\it Axion miniclusters and Bose stars}, Phys. Rev. Lett., {\bf 71}, 3051 (1993).
\bibitem{paper:caustic_rings} P. Sikivie, {\it Caustic rings of dark matter}, Phys. Lett. B, {\bf 432}, 139 (1998).
\end{thebibliography}
\end{document}